%% file: amram.tex
\documentclass[vecphys]{svmult}

\usepackage{makeidx}         % allows index generation
\usepackage{graphicx}        % standard LaTeX graphics tool
\usepackage{multicol}        % used for the two-column index
\usepackage[bottom]{footmisc}% places footnotes at page bottom
\makeindex             % used for the subject index
\newcommand {\ha} {H$\alpha$\,\,}

\newcommand {\kms} {\,km\,s$^{-1}$\,}

\newcommand{\Arc}{''}

\begin{document}

\title*{From Nearby to High Redshift Compact Group of Galaxies. }
% \titlerunning{Short Title} for an abbreviated version of
\author{Philippe Amram\inst{1}\and
Chantal Balkowski\inst{2}\and Claudia Mendes de Oliveira\inst{3}\and Henri Plana\inst{4}\and Beno\^{i}t Epinat\inst{1}}
\authorrunning{Amram et al.}
\institute{Laboratoire d'Astrophysique de Marseille, OAMP, Universit\'{e} de Provence, CNRS, France -- \texttt{Philippe.Amram@oamp.fr}
\and GEPI, Observatoire de Paris-Meudon, Universit\'{e} Paris VII, CNRS, France  \and IAG, Universidade de S\~{a}o Paulo, Brazil\and
Universidade Estadual de Santa Cruz, Ilh\'{e}us, Brazil}
\maketitle

\section{Introduction}
\label{sec:1}
% Always give a unique label
% and use \ref{<label>} for cross-references
% and \cite{<label>} for bibliographic references
% use \sectionmark{}
% to alter or adjust the section heading in the running head
Groups of galaxies are small systems containing a few $L_{*}$ galaxies.  Up to half of all nearby galaxies are in groups or clusters
\cite{geller-huchra83} and more than 50\% of the nearby structure in the universe lies in groups formed by 3 to 20 members
\cite{Tully87}.  Nevertheless, only very few of them  are compact.  Compact Groups of Galaxies (CGs) are isolated entities
containing 4 to 7 galaxies close to one another:  their mean galaxy-galaxy angular separation is of the same order as the diameters
of the galaxies and their velocity dispersions are rather low ($\sim200~km~s^{-1}$).  Most of CGs are physically bound by gravity as
it can be attested by the high fraction of interacting members, they evolve through dynamical friction and may finally merge to form
one single galaxy, called a fossil group (\cite{Vikhlinin99},\cite{Jones03}).

Studies on CGs may be classified in four different epochs: 1) the first ones were those of "pioneers", namely Stephan and Seyfert who
discovered respectively Stephan's Quintet (1877) \cite{Stephan1877} and the Seyfert's Sextet \cite{Seyfert48}; the second period
was based on visual inspection on the POSS-I and allowed the identification of numerous pairs, triplets, quartets and quintets of
galaxies (\cite{Vorontsov-Velyaminov57}, \cite{Vorontsov-Velyaminov77}, \cite{Arp66} and \cite{Shakhbasian73}); the third epoch
started with the construction on objectively defined catalogs (\cite{Shakhbasian73},\cite{Rose77},\cite{Hickson82},\cite{Hickson93}),
but still using visual inspection as the search method; and finally the present time, when the searches are mostly done on digital sky
surveys. Digital sky surveys are used with two different methodologies. The first approach consists of using the today large galaxy
redshift surveys (usually surveys done using fiber spectrographs) for identifying CGs as by-products of the catalogues using the  "Las
Campanas Redshift Survey (LCRS)" (\cite{Allam99},\cite{Tucker00}); using the SLOAN \cite{Lee04} or the 2dF galaxy redshift survey
\cite{Merchan-Zandivarez02} . Due to technical limitations imposed by the minimum separation between two adjacent optical fibers
($\sim1'$), the efficiency of this approach is rather low because most galaxies within CGs have angular separation smaller than $1'$.
For instance, in a work based on the LCRS, among 76 candidate CGs only one has measured redshifts for the group; 23 have measures for
2 galaxies; 52 for one galaxy.
Thus, this approach is most suitable to search for loose groups \cite{Merchan-Zandivarez02}.  The second
approach consists of an automatic search on the DPOSS followed by specific redshift determination programs; 122 CGs were discovered
\cite{Prandoni-Iovino-MacGillivray94}; 84 small and high density CGs \cite{Iovino02} and \cite{deCarvalho05} (\cite{deCarvalho05}
include in their sample that of \cite{Iovino02}).

\section{The Stephan's Quintet}

It is a challenge to recover the history of this nearby CG. This group clearly illustrates the complexity of the formation and
evolution of CGs. The Stephan's Quintet was discovered by Edouard Stephan in 1877, using the Foucault 80-cm reflector in Marseille. It
consists of 5 members; three of them are strongly interacting, the fourth one is an elliptical galaxy to the southwest of the group
and the fifth one, NGC7320 is a foreground galaxy, with a velocity of about 800 km/s. Three of the members have about the same
redshift (V(NGC7317) = 6563 \kms; V(NGC7318A) = 6620 \kms and V(NGC7319) = 6650 \kms) and there is a fourth concordant redshift galaxy
(V(NGC 7320C) = 6000 \kms), that was not originally in the quintet, but which is definitely part of the system. It is located 3' from
the other three objects, and it is thought to be a past intruder which produced an old tidal tail \cite{Arp73} embedded in a very
extended HI tail \cite{Williams02}. It is noticeable that there is no detected neutral hydrogen in any galaxy of the CG,  while a mass
of 10$^{10}$ M$_\odot$ of HI is detected in several tails distributed in the intragroup medium. A large cloud of molecular gas
detected through $^{12}CO$ is located at the end of one optical tail, in the middle of the eastern HI tail (BIMA observations
\cite{Gao-Xu00}; IRAM observations \cite{Lisenfeld04}). This large cloud might lead to the formation of a tidal dwarf galaxy
candidate. Seven other tidal dwarf candidates have been detected within an optical tail. They were classified as tidal dwarf
candidates from the study of their kinematic properties obtained from \ha emission \cite{Mendes-de-Oliveira01}. Four intergalactic HII
regions have also be detected in (or closed to) the end of the HI tail, 25 kpc away from the closest bright galaxy member
\cite{Mendes-de-Oliveira04}. A plausible scenario explaining the recent history of this CG may be the following \cite{Sulentic01}: NGC
7318B is entering the group with a different velocity of about 1000 \kms (V(NGC7318B)=5770 \kms) and produces a new optical tidal tail
and particularly a large scale shock region revealed by \emph{(i)} a 40 kpc non thermal radio continuum region
(\cite{van_der_Hulst-Rots81}, \cite{Xu03}); \emph{(ii)} the exceptional detection of the molecule H$_2$ in the IR \cite{Appleton06};
\emph{(iii)} the high resolution X-ray emission detected with Chandra \cite{Trinchieri03} coincident with the \ha emission regions
\cite{Plana99} and the gas stripped regions detected with XMM \cite{Trinchieri05}.  Moreover, the GALEX data in the far and near-UV
detected around NGC 7319 and NGC 7318b show a peculiar morphology but a SFR consistent with that of normal Sbc galaxies, meaning that
the strength of star formation activity is not enhanced by interactions \cite{Xu05}.

\section{Interaction and Merging in Compact Groups}

One of the best pieces of evidence showing that CGs are physically bound is the high rate of groups which are X-ray loud from where it
has been inferred that 75\% of the Hickson compact group sample contains X-rays \cite{Ponman96}. The X-ray emittig CGs are in most
cases dominated by an elliptical galaxy while CGs dominated by spirals do not show X-ray emission \cite{Zabludoff-Mulchay98}.
Commonly, the higher activity is found in CGs having the lowest velocity dispersions. In addition, CGs show a strong velocity
dispersion-morphology correlation but no density-morphology correlation. Interactions that do not disrupt galaxies are common in CGs
but mergers occur seldom (\cite{Mendes-de-Oliveira94}, \cite{Mendes-de-Oliveira98}, \cite{Amram03}, \cite{Plana03}). The interactions
in CGs are revealed by the presence of tidal tails (in which tidal dwarf galaxy candidates are found), by their high IR luminosity,
central activity, diffuse intergalactic light, kinematics and dynamics, HI deficiency and central double nuclei.  For instance, the
diffuse light in the Seyfert's sextet (HCG 79) forms about 46\% of the total light, its blue color indicates that the light may come
from the stripping of dwarf galaxies (bluer than the galaxies) dissolved into the group potential well \cite{daRocha05}. Some
apparently rather quite normal-looking CGs show galaxies with almost all the interaction indicators. This is the case for the galaxy
HCG 16c who displays a highly disturbed velocity field, a central double nuclei, a double kinematic gas component, a kinematic warp, a
gaseous versus stellar major-axis misalignment, a high IR luminosity and central activity \cite{Mendes-de-Oliveira98}. The HI content
in CGs is still somewhat controversial: HI deficiency was found in CGs by one team \cite{Verdes-Montenegro01} while another one did
not find it in CGs while deficiency is found in loose groups \cite{Stevens04}. On the other hand, observational evidences show that
CGs evolution is different from interacting pairs of galaxies: interactions in pairs trigger an inflow of gas to the galactic nucleus
but CGs do not appear to show rates of either activity enhanced beyond that of the field (\cite{Pildis95}, \cite{Allam99}).
Furthermore, traces of mergers are rare in CGs ($\sim$7\%, \cite{Zepf93}). Nevertheless the expected lifetime is much longer than of a
few crossing times. Indeed, CGs have high density and low velocity dispersion so they should merge fast, instead of that their
observed merging rate is very low. Why CGs dominated by ellipticals are still observed today ? Stellar populations of early-type
galaxies in CGs are older than in the field and similar to cluster \cite{Proctor94}. This indicates that despite the fact that
ellipticals in CGs were formed long ago, the CGs have not yet merge.  The crossing time in a CG is a few 0.01 $H_0^{-1}$ while the
galaxy/galaxy merging time is only a few crossing times (determined by n-body simulations), then major merging should be faster than
one Gyr, which is not the case. Two scenarii have been developed to explain the longevity of CGs. The first one needs a continuous
secondary infall of close-by galaxies coming from surrounding loose groups \cite{Ramella94} and the second one invokes either specific
initial condition or massive halos to stabilize CGs (\cite{Athanassoula-Makino-Bosma97}, \cite{Athanassoula00},
\cite{Gomez-Flechoso_Dominguez-Tenreiro01}) allowing to reach dynamical timescales as high as 25 Gyr. Finally, a well known
cosmological issue of the $\Lambda CDM$ theory predicts more satellite galaxies than seen in groups (\cite{Moore99}, \cite{Kyplin99}).

\section{Past and Future of Compact Groups}

Very little is known on the precursors and the end products of CGs. When extreme cases of dynamical friction occur, fossil groups may
be the end products of merging of $L_*$ galaxies in low-density environments \cite{Jones03}. The most massive versions of today's CGs,
like HCG 62, are the best candidate precursors of fossil groups \cite{Mendes-de-oliveira05}. High velocity dispersion CGs, which
contain mainly bright ellipticals and extended X-ray halos, embedded in neighborhoods rich in low-luminosity companions may have been
the precursors of fossil groups \cite{Mendes-de-Oliveira06}. Due to their high sizes and masses, most of the 15 fossil groups
candidate discovered up to now may be in fact poor clusters end-products rather than end-products of CGs. The past history of CGs and
their role in galaxy evolution is not better known. Do galaxies in CGs mimic the early-universe galaxies ?  The high densities of CGs
suggests that it might be the case. Nevertheless, local CGs are long-live structures with a low merging rate even if the interaction
rate between their galaxies is high. What is the fraction of galaxies which are in CGs through the age ? Local CGs in the nearby
universe are isolated and due to the highest density of the universe in the past because the universe was denser, isolated CGs should
not have been frequent in the past. The relation between their global properties and the formation and evolution of their galaxies
members may be a clue for their understanding. Further searches for high-z CGs will allow better understanding about the evolution of
the fraction of CGs and will give the opportunity to learn about the change in their average physical properties (mass, velocity
dispersion, radius, star formation rate). Up to z$\sim$0.12 (1.56 Gyr look back time) no change is seen in their properties
\cite{Pompei06}. A related topic of high interest for the knowledge of the role of CGs in the past is the accretion of CGs by clusters
which may be a preprocessing step in galaxy evolution during the high redshift cluster assembly phase. Indeed, $\Lambda CDM$ theory
indicates that galaxy clusters grow by accreting small groups of galaxies falling in along large scale filaments. A blue CG, with a
low velocity dispersion ($\sim$150 \kms), infalling at high speed ($\sim$1700 \kms) into the dynamically young cluster Abell 1367 was
recently discovered (\cite{Iglesias-Paramo02}, \cite{Sakai02}). This example might reproduce, at the present epoch, the physical
conditions that were likely to exist in clusters under formation. Tidal interactions produced extended tidal tail ($\sim$150 kpc) as
well as stellar shells and weakened the potential wells of the CG, making it easier for ram pressure to strip the galaxies ISM. Tidal
forces and ram-pressure fragment this CG and blown out gas and stars through the cluster. Meanwhile, at least ten tidal dwarfs and
extragalactic compact HII regions were formed.  This produced an unusually high \ha emission associated with the two giant galaxies
and made this CG the highest density of star formation activity ever observed in the local clusters \cite{Cortese06}.

\section{Kinematic of High Redshift Galaxies}

The forthcoming studies on high-z galaxies in CGs will aim to quantify the effects of galaxy evolution. Due to their large distances,
high-z galaxies are unfortunately not observable with the same spatial sampling as low-z galaxies. Thus, first of all, it is necessary
to disentangle distance effects from evolution ones. In order to analyze the kinematics of high-z galaxies, control samples of nearby
galaxies, with well studied kinematics, are necessary. Nearby samples of galaxies having a broad range of luminosities/masses,
morphological types, and environments provide a wide range of kinematical signatures (\cite{Garrido05}, \cite{Hernandez05}). Due to
the loss of spatial resolution, these signatures are smoothed for high-z galaxies. Spatial resampling of nearby galaxies can be used
to simulate distant galaxies \cite{Flores06}. The comparison between "resampled" and distant galaxies can help identifying signatures
of mergers, internal features such as bars or even the disk/halo density distributions. A first order use of kinematical analysis is
the study of the evolution of the Tully-Fisher relation which only requests the maximum velocity of the rotation curve. Nevertheless,
a ``believable'' rotation curve may be obtained from a 2D velocity field if the ratio between the optical radius of the galaxy and the
spatial resolution (called B hereafter) is greater or equal to 10 (B must be larger than 7 in the HI \cite{Bosma78}). Thanks to the
advent of adaptative optics, leading to a resolution of typically 0.1\Arc, B$\geq$10, for a galaxy having a size $\geq2$\Arc. Thus,
the determination of the kinematical parameters of the galaxy such as the dynamical center, its inclination, position angle and its
maximum rotational velocity ($V_{max}$) become reliable. When B is large enough, $V_{max}$ may be computed from the rotation curve
rather than from the width of the central velocity dispersion, in the center of the galaxy.

Even at low redshift, controversy may exist on the nature and on the history of CGs.  It is the case for the nearby HI and \ha gas
rich system HCG 31, displaying a low velocity dispersion ($\sim$60 \kms) and an intense star formation rate. Three scenarios have been
put forward to explain the nature of this CG: \emph{(i)} these are two systems that are in a pre-merger phase (\cite{Amram04},
\cite{Verdes-Montenegro05}), \emph{(ii)} the system is a late-stage merger \cite{Williams91} or \emph{(iii)} it is a single
interacting galaxy \cite{Richer03}.  What would the observations of such a CG tell us if observed at higher redshift ? To tackle this
question, we have tested the beam smearing effect on it.  The original data cube has been convolved by the mean seeing function
(0.62") to simulate a VLT/FLAME/GIRAFFE observation (pixel size=0.52").  On Fig. 1, the group has been projected to redshift z=0.15
and z=0.60, leading respectively to scales of 2.6 and 6.7 kpc per arcsec.  At z=0.013 (the actual distance of HCG 31), the kinematical
signature of two rotating disks (HCG 31 A and C) producing tidal tails is maybe seen in addition to a very inclined galaxy on the SW
side (HCG 31B). At z=0.15, it becomes difficult to count how many galaxies are involved in the system, nevertheless it is still
possible to distinguish at least two galaxies (A+C and B).  At z=0.60, disentangling the system is a real challenge.  At z=0.013, high
spatial and spectral Fabry-Perot observations allow to observe that the broader \ha profiles (larger than 30 $km~s^{-1}$) are located
in the overlapping regions between the two galaxies (HCG 31 A and C). This clearly maps the shock between both galaxies and the
subsequent starburst regions \cite{Amram04}.  At higher redshift, even at z=0.15, this broadening would be interpreted as a indicator
of rotating disk and this system should have been catalogued like a rotator instead of a merger \cite{Flores06}.

On the other hand, in cases when B$\geq$10, it becomes possible to address the problem of the shape of the inner density profile in
spirals (CORE vs CUSPY controversy), which remains one of the five main further challenges to $\Lambda$CDM theory
\cite{PrimackAstro-ph0609541}. For the largest and brightest galaxies, it will then be possible to trace the mass distribution in
separating luminous from dark halo contribution, which has not been done so far; for instance, for the best data observed without
adaptative optics (Q2343-BX610 located at $z\sim2.2$), B$<$3 \cite{Forster-Schreiber06}.

In order to test the systematics induced by the beam smearing effects, we have projected the data cube of the galaxies used to study
the local Tully-Fisher (TF) relation for CGs \cite{Mendes-de-Oliveira03} to different redshifts. The results indicate that the high-z
galaxies have smoother rotation curves than the ones for the local galaxies. In other words, a "solid-bodyfication" of the rotation
curve is observed.  The consequence is that there is no indication that the maximum velocity of the rotation curve is reached, leading
to uncertainties in the TF relation determination.  In addition, the (fine) structures within the galaxies (bars, rings, spiral arms,
bubbles, etc.) are attenuated or erased and the determination of the other kinematical parameters (position angle, center,
inclination) becomes highly uncertain. For the sample we used \cite{Mendes-de-Oliveira03}, we have determined that, up to $z\sim0.6$,
the maximum rotation velocity reached in the rotation curve decreases with redshift, following the empirically derived function:
\begin{equation}\label{titi}
    \frac{\Delta~V}{V}=\frac{V^{max}_{z}-V^{max}_{z=0}}{V^{max}_{z=0}}=-(11\pm4)~10^{-3}z
\end{equation}
For instance, at $z\sim0.6$, a galaxy having an actual rotation velocity $V^{max}$=200 \kms, might be observed with
$V^{max}$=~187$\pm$5 \kms, shifting the TF relation determination towards higher luminosity.  This estimate is a higher limit for
$V^{max}$, since the simulation has been done using the kinematical parameters determined for $z=0$. Indeed, a wrong determination of
the position angle of the major axis will lower $V^{max}$ even more. In addition to this systematic effect, a wrong determination of
the inclination of the disk increases the dispersion of $V^{max}$. Indeed, inclination is the most complicated kinematical parameter
to determine for high-z galaxies due to beam smearing effect and morphological evolution in galaxies. A disk rotating at $V^{max}$=200
\kms inclined to $35^\circ$ with respect to the plane of the sky, might be confused with a disk rotating at $V^{max}$=~160 or ~270
\kms if the inclination is respectively overestimated by 10\% ($25^\circ$) or underestimated by 10\% ($45^\circ$).

\section{Summary/Conclusions}
\begin{itemize}
    \item Nearby Compact Groups of Galaxies (CGs) are very complex systems, tracing their history is a challenge (e.g. Stephan's Quintet).
    \item The presence of a diffuse X-rays emission that often peaks in the center of CGs shows that CGs are bound structures,
they show numerous signs of interaction but their lifetime in much longer than their crossing
    times.
    \item Hickson CGs clearly show different stages of evolution, from weakly interacting galaxies to merging systems.
    \item CGs infalling into clusters may provide a mechanism to form clusters at high redshifts (e.g. in A1367).
    \item Massive versions of today's CGs may have been the best candidate precursors of fossil groups.
    \item Do CGs mimic the high redshift universe? This is still an open question. Indeed, their high density and low velocity dispersion
should induce a high interaction rate and fast merging, CGs are nevertheless long-lived structures. On the other hand, there is
probably no (or a few) isolated CGs in the high z universe.
CGs may fuel high z clusters.   CGs may produce fossil groups and
fossil ellipticals.
    \item CGs at high z are difficult to detect and are still to be discovered.
    \item Interpretation of distant kinematics of galaxies may need nearby sample of galaxies to disentangle beam-smearing from
evolutionary effects (e.g. HCG 31)
    \item Beam smearing effects may bias the Tully-Fisher relation (shifted towards lower M/L)
\end{itemize}

%Your text goes here. Use the \LaTeX\ automatism for your citations \cite{claudia}.

%\subsection{Subsection Heading}
%\label{sec:2}
%Your text goes here.

\begin{figure}
\centering
\includegraphics[height=4cm]{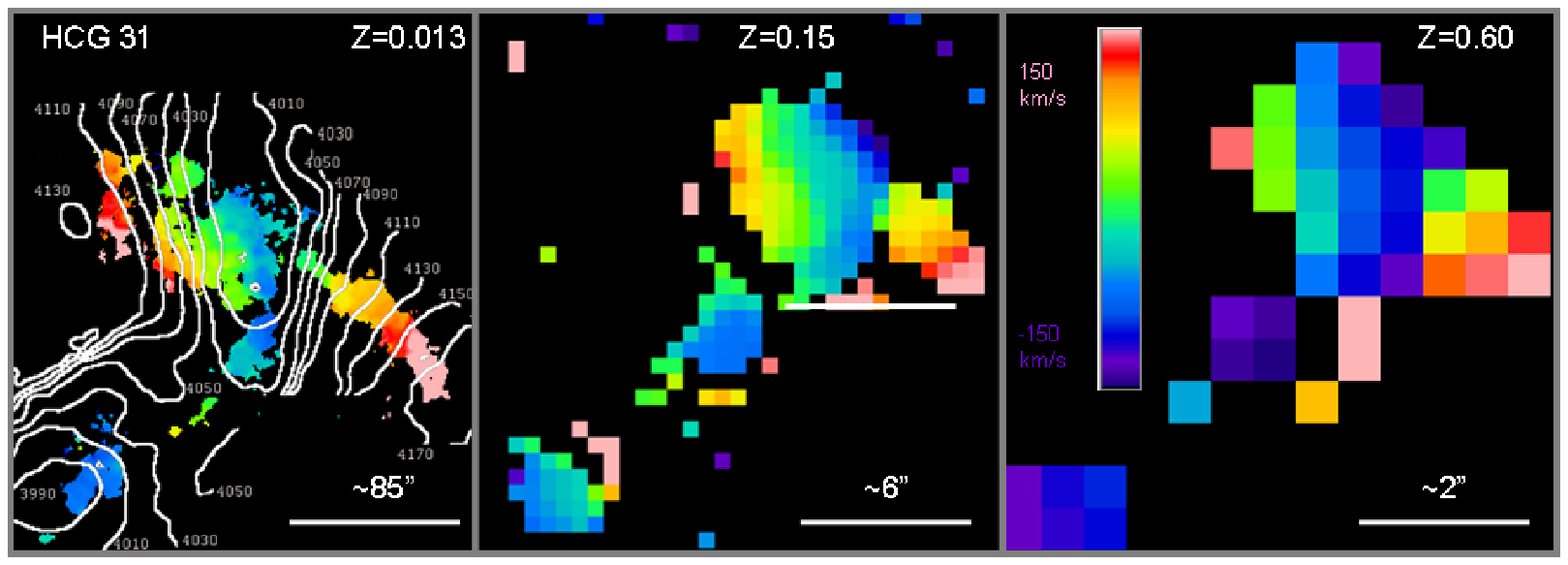}
%\picplace{5cm}{2cm} % Give the correct figure height and width in cm
%
\caption{-Left- \ha Velocity Field of HCG 31 on which is superimposed the HI velocity contours from Verdes-Montenegro et al. - Middle
and right - The original data cube of HCG 31 has been convolved by the mean seeing function (0,62") to simulate a VLT/FLAME/GIRAFFE
observation (pixel size=0.52"). The group has been projected respectively to redshift z=0.15 (middle image) and z=0.60 (right image),
leading respectively to a scale of 2.6 and 6.7 kpc per arcsec ($h_0=71$, $\Omega_M=0.27$, $\Omega_V=0.73$). The velocity scale is the
same for the 3 images [64].}
% \cite{Amram04}.}
\label{fig:1}       % Give a unique label
\end{figure}
%
%
% BibTeX users please use
% \bibliographystyle{}
% \bibliography{}
%
% Non-BibTeX users please follow the syntax
% the syntax of "referenc.tex" for your own citations
\input{amram_referenc}

%%%%%%%%%%%%%%%%%%%%%%%%%%%%%%%%%%%%%%%%%%%%%%%%%%%%%%%%%%%%%%%%%%%%%%  }

%%%%%%%%%%%%%%%%%%%%%%%%%%%%%%%%%%%%%%%%%%%%%%%%%%%%%%%%%%%%%%%%%%%%%%

\printindex
\end{document}

%% file: amram_referenc.tex
%%%%%%%%%%%%%%%%%%%%%%%% referenc.tex %%%%%%%%%%%%%%%%%%%%%%%%%%%%%%
% sample references
% "physics"
%
% Use this file as a template for your own input.
%
%%%%%%%%%%%%%%%%%%%%%%%% Springer-Verlag %%%%%%%%%%%%%%%%%%%%%%%%%%

%
% BibTeX users please use
% \bibliographystyle{}
% \bibliography{}